\documentclass[%
 reprint,
 superscriptaddress,
 preprintnumbers,
 twocolumn,
nofootinbib,
 amsmath,amssymb,
 aps,
prd,
 floatfix,subfloat
]{revtex4}
\usepackage{verbatim}
\usepackage{multirow}
\usepackage{graphicx}
\usepackage{dcolumn}
\usepackage{bm}
\usepackage{hyperref}
\usepackage{url}
\usepackage{cleveref}

\usepackage{physics}
\usepackage{color}
\DeclareUnicodeCharacter{2212}{-}

\begin{document}

\preprint{ICPP-057, PSI-PR-21-21, ZU-TH 38/21, CERN-TH-2021-129, LTH 1267}

\title{Accumulating Evidence for the Associated Production of a \\ New Higgs Boson at the Large Hadron Collider}

\author{Andreas Crivellin}
\email{andreas.crivellin@cern.ch}
\affiliation{Physik-Institut, Universität Zürich, Winterthurerstrasse 190, CH–8057 Zürich, Switzerland}
\affiliation{Paul Scherrer Institut, CH–5232 Villigen PSI, Switzerland}
\affiliation{CERN Theory Division, CH–1211 Geneva 23, Switzerland }

\author{Yaquan Fang}
\email{fang@mail.cern.ch}
\affiliation{Institute of High Energy Physics, 19B, Yuquan Road, Shijing District, Beijing, China, 100049}
\affiliation{University of Chinese Academy of Sciences (CAS), 19A Yuquan Road, Shijing District, Beijing, China, 100049}
\author{Oliver Fischer}
\email{oliver.fischer@liverpool.ac.uk}
\affiliation{Department of Mathematical Sciences, University of Liverpool, Liverpool, L69 7ZL, UK}

\author{Srimoy Bhattacharya}
\email{bhattacharyasrimoy@gmail.com}
\affiliation{School of Physics and Institute for Collider Particle Physics, University of the Witwatersrand,
Johannesburg, Wits 2050, South Africa.}

\author{Mukesh Kumar}
\email{mukesh.kumar@cern.ch}
\affiliation{School of Physics and Institute for Collider Particle Physics, University of the Witwatersrand,
Johannesburg, Wits 2050, South Africa.}

\author{Elias Malwa}
\email{elias.malwa@cern.ch}
\affiliation{School of Physics and Institute for Collider Particle Physics, University of the Witwatersrand,
Johannesburg, Wits 2050, South Africa.}
\affiliation{
iThemba LABS, National Research Foundation, PO Box 722, Somerset West 7129, South Africa.}

\author{Bruce Mellado}
\email{bmellado@mail.cern.ch}
\affiliation{School of Physics and Institute for Collider Particle Physics, University of the Witwatersrand,
Johannesburg, Wits 2050, South Africa.}
\affiliation{
iThemba LABS, National Research Foundation, PO Box 722, Somerset West 7129, South Africa.}

\author{Ntsoko Rapheeha}
\email{ntsoko.phuti.rapheeha@cern.ch}
\affiliation{School of Physics and Institute for Collider Particle Physics, University of the Witwatersrand,
Johannesburg, Wits 2050, South Africa.}
\affiliation{
iThemba LABS, National Research Foundation, PO Box 722, Somerset West 7129, South Africa.}

\author{Xifeng Ruan}
\email{xifeng.ruan@cern.ch}
\affiliation{School of Physics and Institute for Collider Particle Physics, University of the Witwatersrand,
Johannesburg, Wits 2050, South Africa.}

\author{Qiyu Sha}
\email{shaqiyu@ihep.ac.cn}
\affiliation{Institute of High Energy Physics, 19B, Yuquan Road, Shijing District, Beijing, China, 100049}
\affiliation{University of Chinese Academy of Sciences (CAS), 19A Yuquan Road, Shijing District, Beijing, China, 100049}

\begin{abstract}
In the last decades, the Standard Model (SM) of particle physics has been extensively tested and confirmed, with the announced discovery of the Higgs boson in 2012 being the last missing puzzle piece. Even though since then the search for new particles and interactions has been further intensified, the experiments ATLAS and CMS at the Large Hadron Collider (LHC) at CERN did not find evidence for the direct production of a new state. However, in recent years deviations between LHC data and SM predictions in multiple observables involving two or more leptons (electrons or muons) have emerged, the so-called ``multi-lepton anomalies'', pointing towards the existence of a beyond the SM Higgs boson $S$. While from these measurements its mass cannot be exactly determined, it is estimated to lay in the range between $130\,$GeV and $160\,$GeV. Motivated by this observation, we perform a search for signatures of $S$, by using existing CMS and ATLAS analyses. Combining channels involving the associate productions of SM gauge bosons ($\gamma\gamma$ and $Z\gamma$), we find that a simplified model with a new scalar with $m_S= 151.5$\,GeV is preferred over the SM hypothesis by 4.3$\sigma$ (3.9$\sigma$) locally (globally). On the face of it, this provides a good indication for the existence of a new scalar resonance $S$ decaying into photons, in association with missing energy and allows for a connection to the long-standing problem of Dark Matter. Furthermore, because $S$ is always produced together with other particles, we postulate the existence of a second new (heavier) Higgs boson $H$ that decays into $S$ and propose novel searches to discover this particle, which can be performed by ATLAS and CMS.
\end{abstract}
\maketitle


\section{Introduction} 
The SM of particle physics describes very successfully the fundamental constituents of matter and their interactions. It has been extensively tested and verified experimentally~\cite{ParticleDataGroup:2020ssz,HFLAV:2019otj,ALEPH:2005ab} within the last decades with the discovery of the Brout-Englert-Higgs boson ($h$)~\cite{Higgs:1964ia,Englert:1964et,Higgs:1964pj,Guralnik:1964eu} at the LHC~\cite{Aad:2012tfa,Chatrchyan:2012ufa} being the last missing puzzle piece. Furthermore, measurements so far indicate that this $125\,$GeV boson has properties compatible with those predicted by the SM~\cite{Chatrchyan:2012jja,Aad:2013xqa}. However, this does not exclude the existence of additional scalar bosons as long as their role in the breaking of the SM gauge symmetry is sufficiently small. In fact, the searches for new particles, including additional Higgs bosons, have been intensified since the Higgs discovery. However, the LHC experiments ATLAS and CMS did not observe the direct (resonant) production of a new particle.
 
Nonetheless, in recent years the so-called ``multi-lepton anomalies" emerged as deviations from the SM predictions in several analyses with multi-lepton (electrons and their heavier version the muons) final states~\cite{vonBuddenbrock:2017gvy,vonBuddenbrock:2019ajh,vonBuddenbrock:2020ter,Hernandez:2019geu}. These intriguing indications for New Physics (NP) are statistically most compelling in non-resonant di-leptons (i.e. lepton not originating from the direct decay of a new particle). These signatures can be explained by the decay of a neutral scalar $H$ (a new Higgs boson) into a lighter new Higgs $S$ and the SM Higgs~\cite{vonBuddenbrock:2016rmr,vonBuddenbrock:2018xar}, i.e. $H\rightarrow Sh,SS$.

The explanation of the multi-lepton anomalies requires the mass of $S$ to be between $130\,$GeV and $160\,$GeV~\cite{vonBuddenbrock:2017gvy}. Fortunately, this mass range is covered in the CMS and ATLAS searches for the SM Higgs. Therefore, the published analyses of di-photon, $Z\gamma$ and $b\overline{b}$ resonances~\cite{Sirunyan:2021ybb,ATLAS:2020pvn,Aad:2020ivc,Sirunyan:2020sum,Aad:2021qks,CMS:2018nlv,Sirunyan:2018tbk,HIGG-2018-51web} can be used to search for a signal of $S$. 
Importantly, it is possible to combine the information from these channels without specifying an explicit model, such that the only assumption is the existence of a scalar particle $S$ which is produced in association with other particles.

In this article we present a combined fit to the available data to search, briefly discuss the implications of the different signals for the properties of $S$, and suggest new searches for $H$ to verify our hypothesis. 

\section{Analysis and Results}

\begin{table*}[t]
    \centering
    \begin{tabular}{|c|c|c|c|}
    \hline
         Channel & Collaboration (category) & (Acceptance $\times$ efficiency) in \% & Resolution  \\
         \hline
          ${S(\to Z(\to \ell^+ \ell^-)\gamma) + \ell }$ & CMS, electron (muon) & 18-24 (25-31) \cite{Sirunyan:2018tbk} & 2.31-4.01 (1.94-3.89)  \cite{CMS:2013rmy}\\
        \hline
        \multirow{2}{*}{$S(\to b\bar{b}) + E^T_{\rm miss}$} &  ATLAS ($qq \to ZH \to \nu\nu b\bar{b} $) & 1.9 \cite{HIGG-2018-51web} & \multirow{2}{*}{13.86 \cite{ATLAS:2014vuz}}\\
          \cline{2-3}
           & ATLAS ($gg \to ZH \to \nu\nu b\bar{b} $) & 3.5 \cite{HIGG-2018-51web} & \\
            \hline
            \multirow{2}{*}{$S(\to \gamma\gamma) + E^T_{\rm miss}$}\footnote{This analysis was originally preformed in the context of a $Z^\prime$ model with $pp\to Z^\prime\to h+{\rm MET}$. We recasted it in the framework of our simplified model with $pp\to H\to SS^*$.}& CMS (High- $E^{\rm miss}_T$) & 42.6 \cite{CMS:2018nlv}                                    &  1.60 - 3.17 \cite{CMS:2018piu}\\
           \cline{2-4}
           & ATLAS (High - $E^{\rm miss}_T$) & 38.7  \cite{ATLAS:2014cnc}                                   & 1.22 -2.61 \cite{ATLAS:2020pvn}\\
        \hline
            \multirow{2}{*}{${S (\to \gamma \gamma) + b}${-jet}} & ATLAS & 42 \cite{ATLAS:2018hxb} & 1.42 - 2.31 \cite{ATLAS:2020pvn}\\
            \cline{2-4}
            & CMS & 40-44  \cite{CMS:2018piu}                                   & 1.64-2.33 \cite{CMS:2021kom}\\
            \hline
            \multirow{2}{*}{$S (\to \gamma \gamma) + W,Z$} & CMS & 79 \cite{CMS:2021kom}& 2.03 - 2.80 \cite{CMS:2021kom}\\
            \cline{2-4}
            & ATLAS & 21.4 \cite{ATLAS:2018hxb} & 1.64 - 2.61 \cite{ATLAS:2020pvn}\\
            \hline
            \multirow{2}{*}{$S\to \gamma \gamma$ (inclusive)} & CMS & 79 \cite{CMS:2021kom} & 1.47 - 3.08 \cite{CMS:2021kom}\\
            \cline{2-4}
            & ATLAS & 21.77 \cite{ATLAS:2022tnm} & 1.22-2.37 \cite{ATLAS:2020pvn}\\
            \hline
            \end{tabular}
    \caption{Acceptance times efficiency and resolution for the invariant mass distribution of $S$ in the different channels given in the ATLAS and CMS analyses. }
    \label{tab:acceptance}
\end{table*}

\begin{table}[t]
\scriptsize
\centering
\begin{tabular}{c|ccc}
\hline\hline
Channel	&	Cross sec. [fb]	&	Obs. lim. &	Exp. lim. \\
$S(\gamma\gamma)$	&	6.6$\pm$3.2	&	14.5	&	8.7	\\
$S(\gamma\gamma)$ + $E^{T}_{miss}>$90\,GeV	&	0.63$\pm$0.20	&	0.98	&	0.33	\\
$S(\gamma\gamma)$+$V\rightarrow jj$ 	&	0.42$\pm$0.42 &	1.20	&	0.86	\\
$S(\gamma\gamma)$ + $b$-jets &	0.12$\pm$0.12	&	0.33 &	0.24	\\

$S(\ell\ell\gamma)$ + $V\rightarrow \ell\nu$ or $\ell\ell$	& 	1.3$\pm$0.7	&	2.8	&	1.7	\\
$S(b\overline{b})$ + 150$<E^{T}_{miss}<$250\,GeV 	&	0.90$\pm$0.79	&	2.2	&	1.6	\\
\hline
$S(4\ell)$ &	0$\pm$0.15	&	0.28	&	0.33 \\
\hline\hline
        \end{tabular}
        \caption{
Extracted cross sections in units of fb for each final state considered (see main text for details). The observed and expected limits on the cross-sections (at 95\% confidence level) are provided. For the second and the last category total cross sections are quoted while for the associate production channels, fiducial ones are given. The $S\to 4\ell$ channel is not included in the fit but rather used as a constraint.}
        \label{tab:yields}
\end{table}

For our analysis we use the sidebands of the SM Higgs analysis of ATLAS and CMS presented in Refs.~\cite{Sirunyan:2021ybb,ATLAS:2020pvn,Aad:2020ivc,Sirunyan:2020sum,Aad:2021qks,CMS:2018nlv,Sirunyan:2018tbk,HIGG-2018-51web}, which cover (implicitly) also the search for other Higgs-like resonances. In fact, depending on the specific channel, searches range up to 180$\,$GeV. However, since some searches only reach $160\,$GeV, we will use the region between 140$\,$GeV (to avoid overlap with SM Higgs signals) and 155\,GeV as the use of a side-band background fit further reduces the range within one can search for a resonance. Note that this coincides with the mass range motivated by the multi-lepton anomalies.

Let us summarize the CMS and ATLAS searches used in our analysis:

{\boldmath$S(Z(\ell^+\ell^-)\gamma) +\ell$}: $m_S$ is reconstructed from the invariant mass of the $Z\gamma$ pair and $S$ is assumed to be produced in association with one additional lepton (other than the leptons originating from the decays of the $Z$). The data are extracted from Fig.~5 in Ref.~\cite{Sirunyan:2018tbk}.
    
{\boldmath$S$($\gamma\gamma$)+$E^{T}_{miss}$}: In these channels $m_S$ is reconstructed from the invariant mass of $\gamma\gamma$ and $S$ is produced in association with $E^{T}_{miss}$. The data are taken from Fig.~6 in Ref.~\cite{Aad:2021qks} and Fig.~3 in Ref.~\cite{CMS:2018nlv}.\footnote{Note that the excess observed in this channel dominantly originates from the ATLAS analysis as ATLAS used four times more luminosity as CMS and the cuts used by ATLAS turn out to be more optimal for this search than the ones of CMS. }
    
{\boldmath$S$($b\overline{b}$)+$E^{T}_{miss}$}: $m_S$ is reconstructed from the invariant mass of $b\overline{b}$ and $S$ is produced in association with $E^{T}_{miss}$ originating from the decay of $S$ to invisible final states.\footnote{It should be mentioned that there is a small overlap between this category and the SM channel $t\overline{t}h$. However, the pollution of the former by the latter is very small (about 1\%).} The data are taken from Fig.~5 in Ref.~\cite{HIGG-2018-51web}.  
    
{\boldmath$S$($\gamma\gamma$)+$b$} jets: $m_S$ decays to two photons and is produced in association with $b$ quarks (which, in the 2HDM+$S$ model, could originate from $S$ but also from $h$ if $H\to Sh$ is non-negligible). The data are obtained from Fig.~2 (top-right) in Ref.~\cite{Aad:2020ivc} and Fig.~2 in Ref.~\cite{Sirunyan:2020sum}.
    
{\boldmath$S$($\gamma\gamma$)+ $V(W,Z)$}: $S$ decays to two photons and is produced in association with a $W$ or a $Z$ boson. The corresponding data are given in Fig.~15 bottom-left in Ref.~\cite{Sirunyan:2021ybb} and Fig.~9 (c) and (d) in Ref.~\cite{ATLAS:2020pvn}.

{\boldmath $S(\gamma\gamma)$ inclusive}: Here $m_S$ is reconstructed from the invariant mass of the photon pair while the search is quasi-inclusive, however, vector boson fusion, $W$ and $Z$ as well as top quark associated production are excluded. Note that there is no veto on missing energy, but that this channel covers only a very tiny phase space of the quasi-inclusive final search.
(see Fig.~15 top-left in Ref.~\cite{Sirunyan:2021ybb} and Fig.~9 (a) in Ref.~\cite{ATLAS:2020pvn}). 

The summary of the values for the resolutions as well as the product of acceptance times efficiency for the different channels is given in Table~\ref{tab:acceptance}.

\begin{figure}[t]
  \centering
   \includegraphics[width=0.45\textwidth]{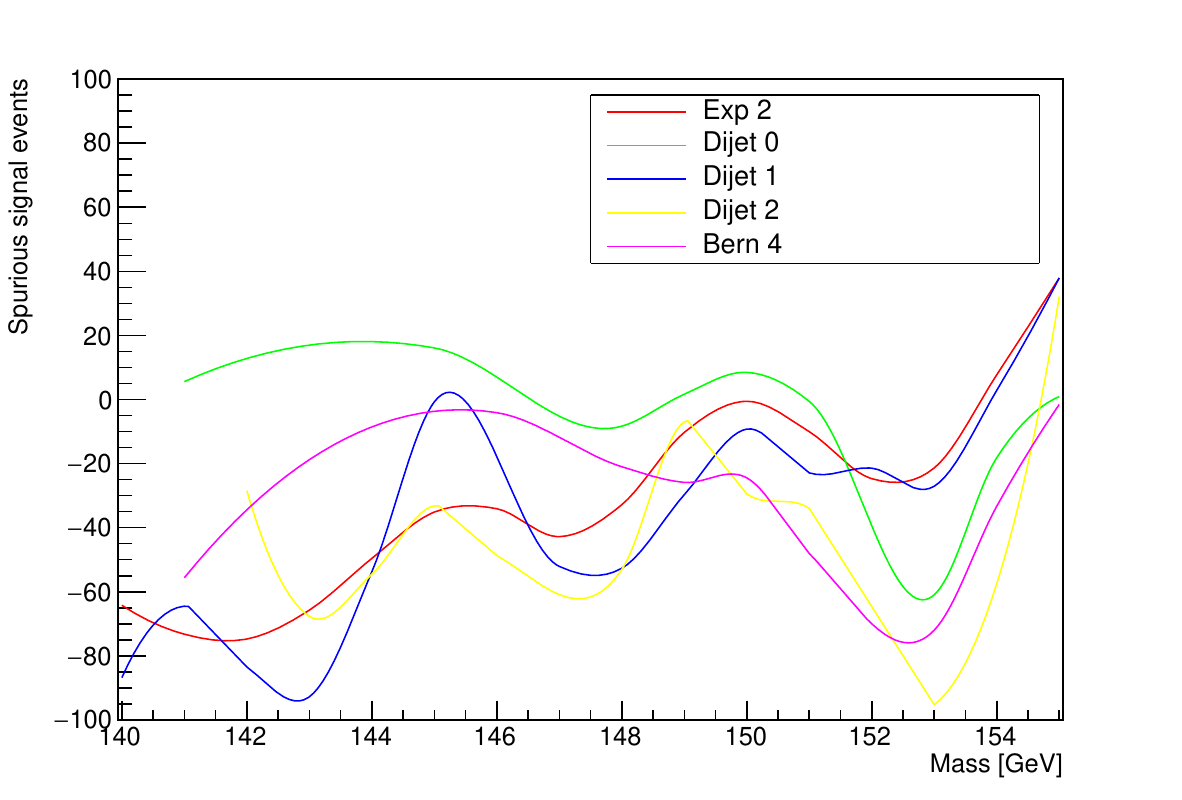}
   \includegraphics[width=0.45\textwidth]{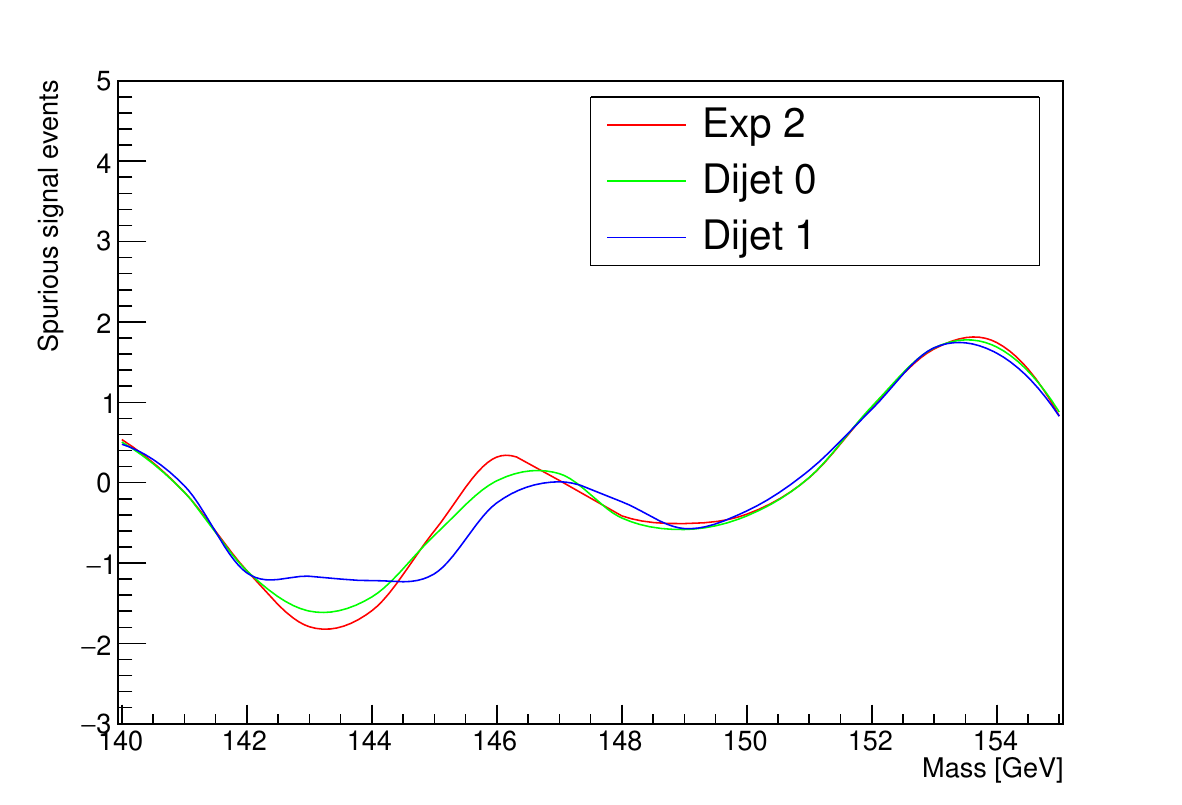}
 \caption{Top: Spurious signal events from a toy MC simulating of inclusive di-photon analysis. Bottom: Spurious signal events from a toy MC simulation of the $E^{T}_{miss}$ category with a threshold on $E^{T}_{miss}$ about 90$\,$GeV. The different functional forms are compared, where Dijet~${1}$ corresponds to Eq.~(\ref{background}). Dijet~${0}$ represents the functional form of Eq.~(\ref{background}) with $a_{1}=0$ and Dijet~${2}$ is obtained from Dijet~${1}$ by adding an additional $log^{2}$ term. Exp~${2}$  stands for the second-order polynomial of the exponential function. Bern~${4}$ is the 4${^{\rm th}}$ order Bernstein function~\cite{bernstein}.  }
    \label{fig:ss}
 \end{figure}

   \begin{figure*}[t]
  \centering
   \includegraphics[width=0.33\textwidth]{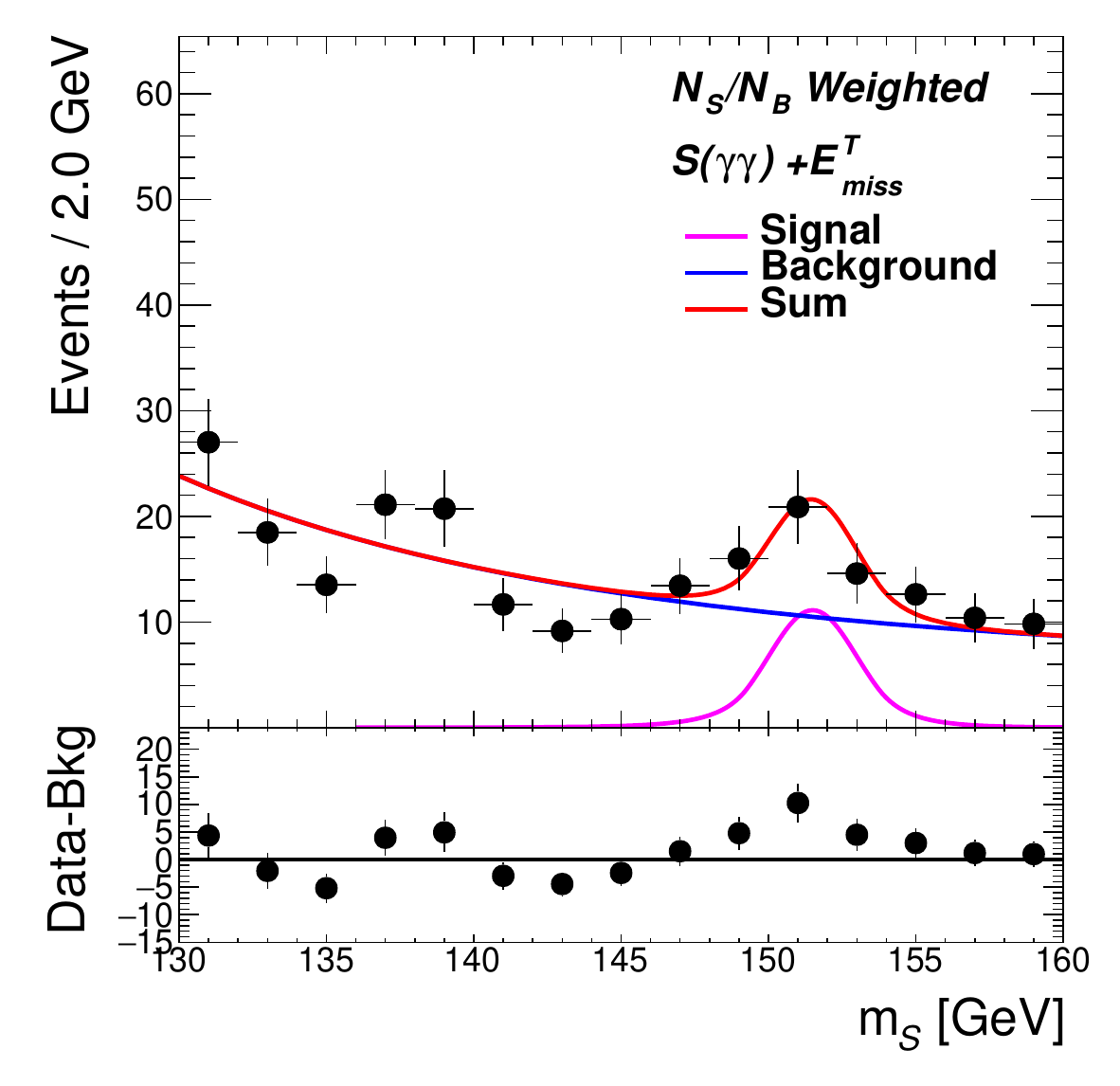}
   \includegraphics[width=0.33\textwidth]{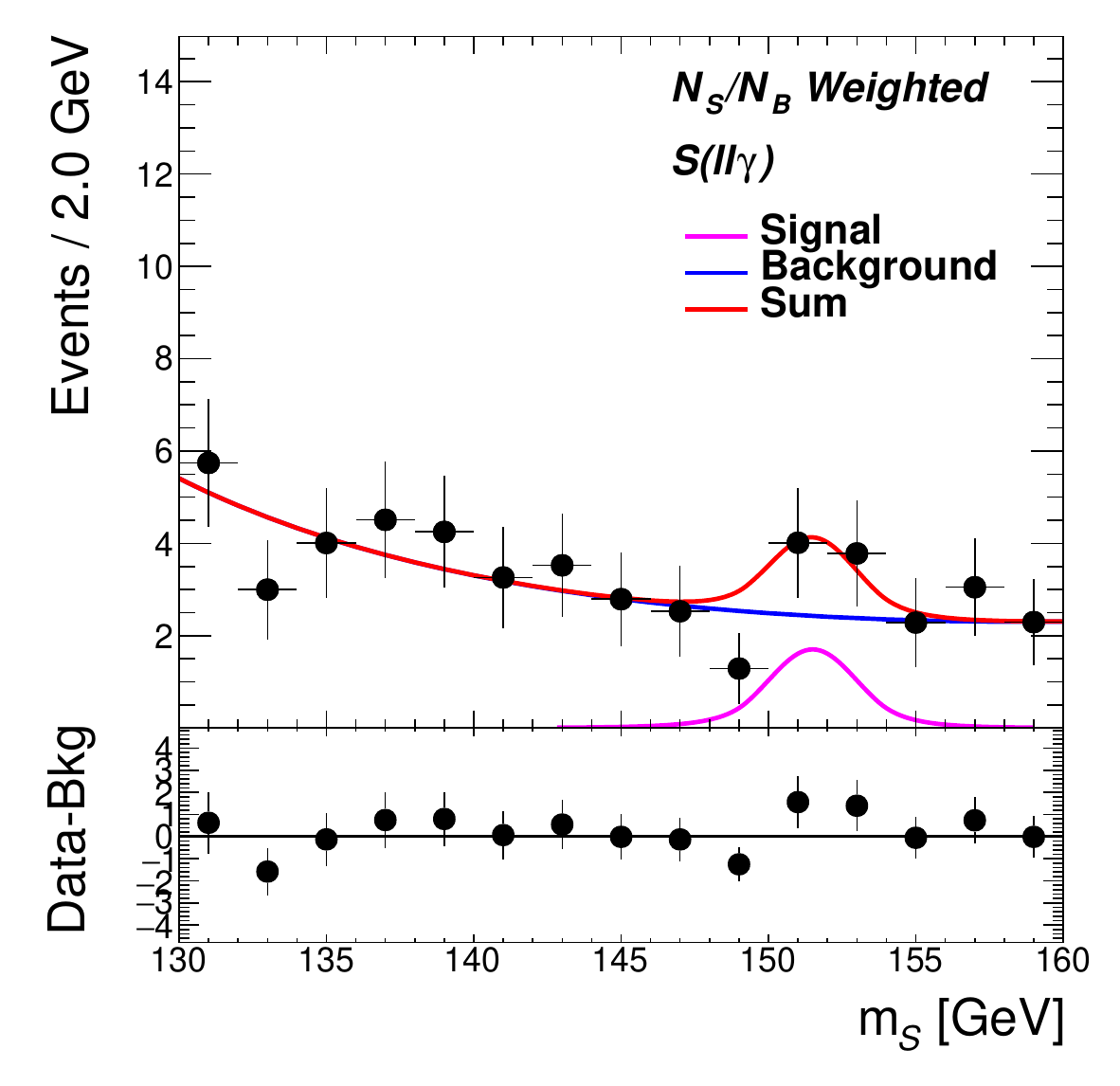}
   \includegraphics[width=0.33\textwidth]{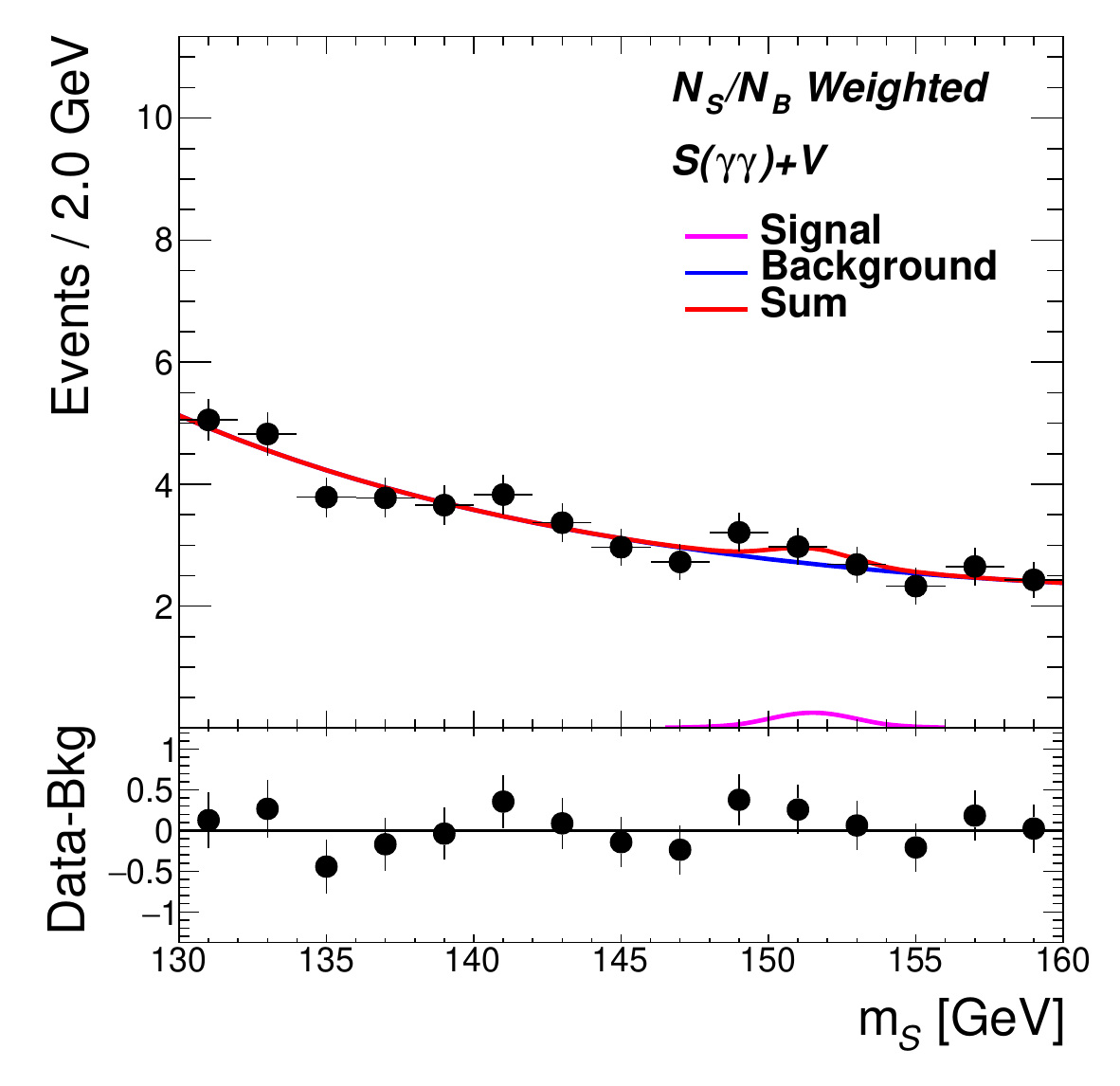}
   \includegraphics[width=0.33\textwidth]{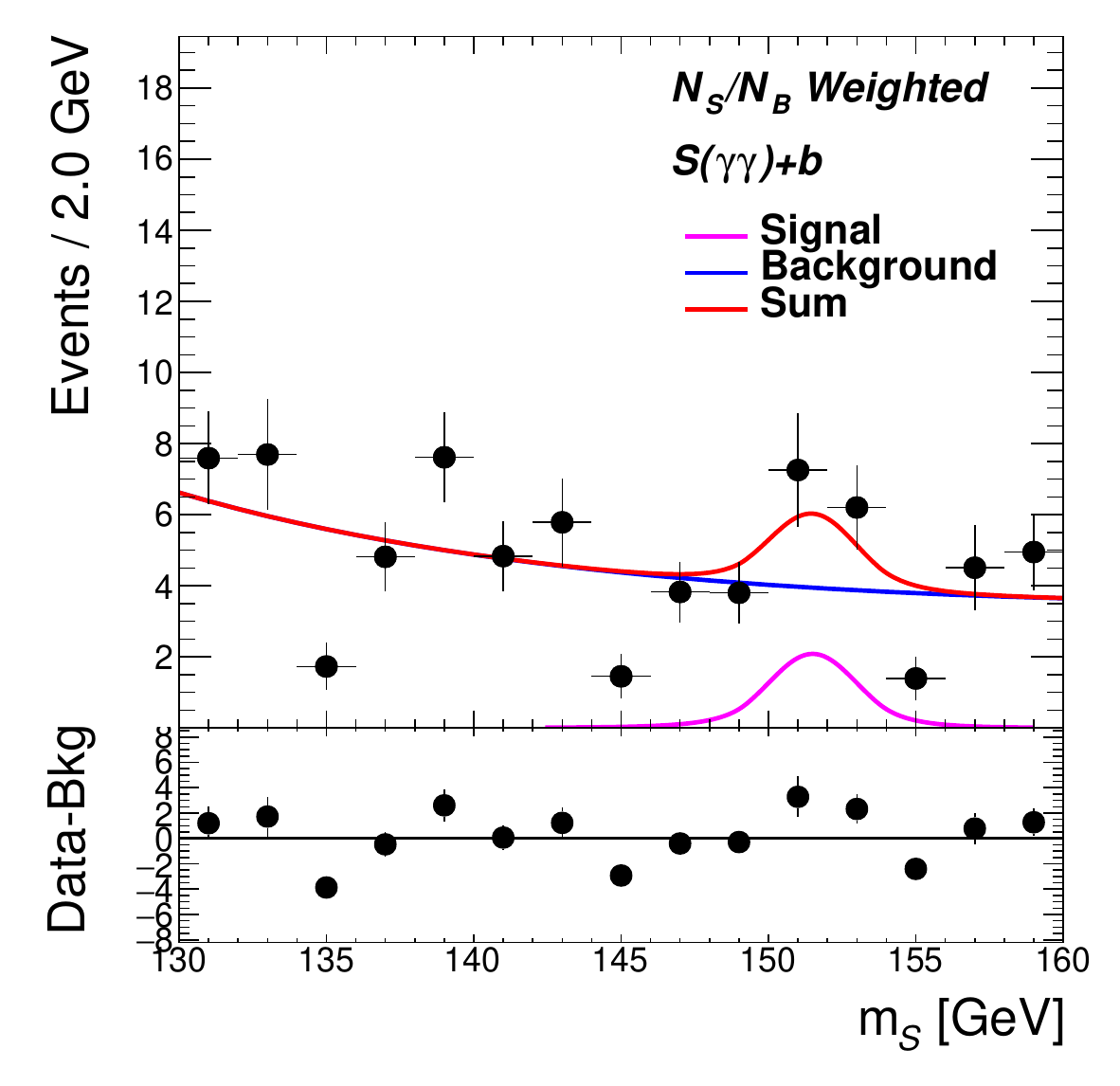}
  \caption{Fits to the spectra of the different categories extracted from Refs.~\cite{Sirunyan:2021ybb,ATLAS:2020pvn,Aad:2020ivc,Sirunyan:2020sum,Aad:2021qks,CMS:2018nlv,Sirunyan:2018tbk,HIGG-2018-51web} (see text for details) obtained from a combination of CMS and ATLAS analyses. The given numbers of events are re-weighted by the ratio of signal over background at $151$ GeV for each channel.  }
    \label{fig:invmass1}
 \end{figure*}
 
To search for a signal in each category, we add to the background the function in Eq.~(\ref{background}) a double-sided-crystal-ball function:
\begin{equation}
   N\cdot 
   \begin{cases}
   e^{-t^{2}/2} & \mbox{if $-\alpha_{\rm Low} \leq t \leq \alpha_{\rm High}$}\\
   \frac{ e^{-0.5\alpha_{\rm Low}^{2} }}{ \left[\frac{\alpha_{\rm Low}}{n_{\rm Low}} \left(\frac{n_{\rm Low}}{\alpha_{\rm Low}} - \alpha_{\rm Low} -t \right)\right]^{n_{\rm Low}} } & \mbox {if $t < -\alpha_{\rm Low} $}\\
   \frac{ e^{-0.5\alpha_{\rm High}^{2} }}{ \left[\frac{\alpha_{High}}{n_{\rm High}} \left(\frac{n_{\rm High}}{\alpha_{\rm High}} - \alpha_{\rm High}  + t \right)\right]^{n_{\rm High}} } & \mbox {if $t > \alpha_{\rm High}$}.
   \end{cases}
   \label{eq:DSCB}
\end{equation}
Here $N$ is a normalization parameter, $t = (m - m_{S})/\sigma_{CB}$ with $\sigma_{CB}$ the width of the Gaussian part of the function, $m$ is the invariant mass of the distribution and $m_S$ the mass of the new resonance we are interested in (which is the same for all channels). $\alpha_{\rm Low}$($\alpha_{\rm High}$) is the point where the Gaussian becomes a power law on the low (high) mass side and we set $\alpha_{\rm Low}=\alpha_{\rm High}=1.5$. $n_{\rm Low}$ ($n_{\rm High}$) is the exponent of this power law and $n_{\rm Low}$ ($n_{\rm High}$) is set to 5 (9). The fit results are quite insensitive to the specific choice of $\alpha_{\rm Low}$($\alpha_{\rm High}$) or $n_{\rm Low}$ ($n_{\rm High}$). As we assume that the physical width of $S$ is much smaller than the detector resolution of the respective channels, $\sigma_{CB}$ is thus determined by the detector resolution:  $\sigma_{CB}=1.5\,$GeV for the di-photon and $Z\gamma$ channel, and $\sigma_{CB}=14\,$GeV for the $b\overline{b}$ channel.

For each category discussed above, we model the background via a function
\begin{equation}
f(m;b,\{a\}) = (1 - m)^b (m)^{a_0 + a_1 \log(m)}\,,
\label{background}
\end{equation}
where $a_{0,1}$ and $b$ are free parameters (different for each category) and $m$ is the invariant mass of the distribution, e.g.~the di-photon mass. This corresponds to the background-only hypothesis and the goodness of the corresponding fit results in the p-value of the SM. The choice of the functional form to model the background is not important for our study, as illustrated by the spurious signals in Fig.~\ref{fig:ss}. For simulating the inclusive di-photon channel, as shown in the top panel, the different functional forms behave very similarly. Within the range of $145\,$GeV-$155\,$GeV, the absolute spurious signal is about 50 events, which corresponds to $\approx 20\%$ of the statistical uncertainty in this channel. Note that all other channels in this study contain much smaller data statistics. For example, for the $E^{T}_{miss}$ channel requiring $E^{T}_{miss}>90\,$GeV, as shown in the bottom panel, the spurious signal is about one event and accounts therefore for about 10\% of the data uncertainty. Moreover, the different choices of the functional form show compatible curves. 

For any given final state (category), the relative contribution of sub-categories to the total signal strength is taken to be proportional to the observed excess in each spectrum. For some of the spectra, this is obtained by a $N_{S}/(N_{S}+N_{B})$ or ${\rm ln}(1+N_{S}/N_{B})$ re-weighting, where $N_{S}$ is the number of SM Higgs events in the sub-category in the original analysis and $N_{B}$ is the number of background events under the Higgs boson peak. The spectra are then unweighted using a constant factor $N^{h}_{\rm unweighted}/N^{h}_{\rm weighted}$, where $N^{h}_{\rm (un)weighted}$ is the number of SM Higgs (125\,)GeV signal evens obtained in the CMS or ATLAS analyses. The simultaneous fit of $m_S$ using all channels therefore takes into account the different integrated luminosity in each analysis. This is then implemented into the likelihood ratio formalism, relying on the software used as well for the discovery of the SM Higgs boson~\cite{Chatrchyan:2012jja,Aad:2013xqa}. 

Table~\ref{tab:yields} shows the extracted total cross-section for the (quasi-inclusive) di-photon analysis, the limit on the total $4\ell$ cross-section (not included in the signal analysis) and the fiducial cross sections (the cross section after including experimental cuts) for all other associated production channels.  
We define the signal yield as $Y = \epsilon \cdot \sigma_{S}\cdot L$, where $\sigma_{S}$ is the total or fiducial cross-section, depending on the analysis, the (channel dependent) luminosity $L$ and the efficiency $\epsilon$, which parametrizes analysis-specific selection criteria and the geometric acceptance of the detector. In Refs.~\cite{ATLAS:2020pvn,Sirunyan:2018tbk}, acceptance times efficiency is $\approx 30\%$. We can roughly estimate the acceptance to be $\approx 50\%$, corresponding to an efficiency of $\approx 60\%$, and therefore the total cross section to be $\approx 2$ times the fiducial one.

\begin{figure*}[t!]
  \centering
     \includegraphics[width=0.54\textwidth]{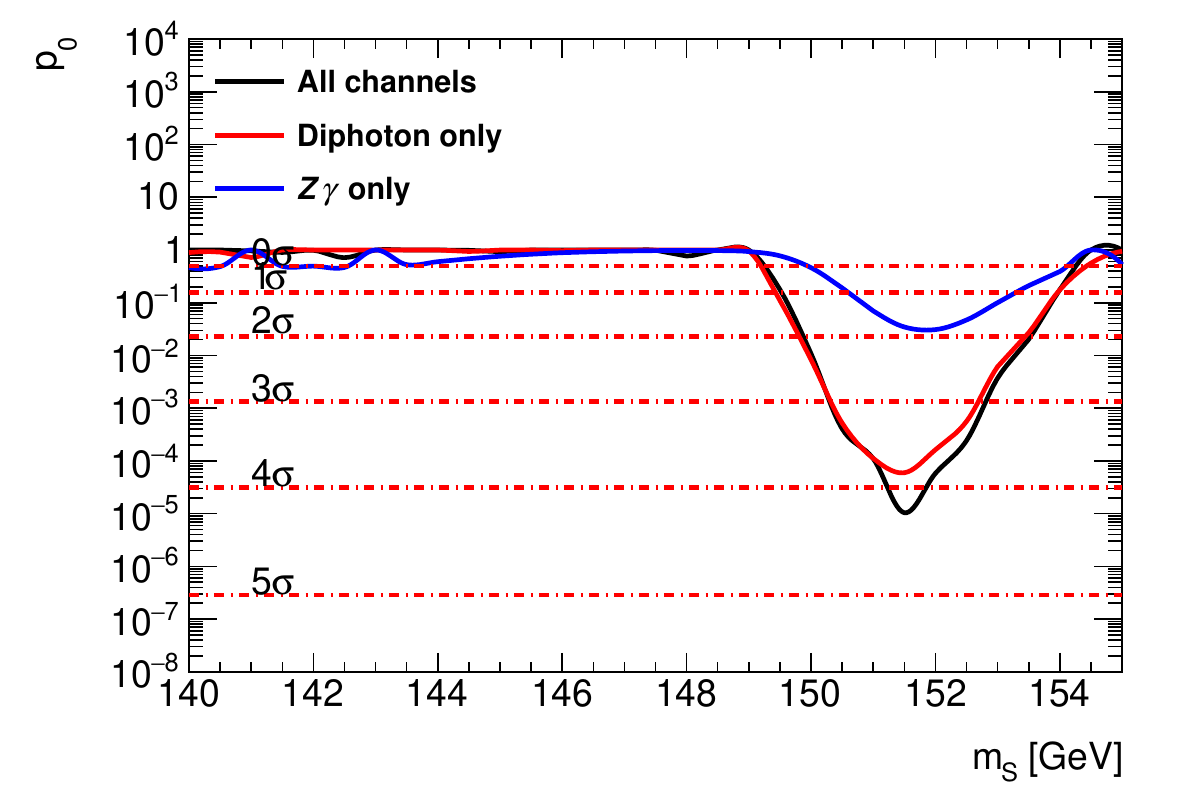}
   \includegraphics[width=0.44\textwidth]{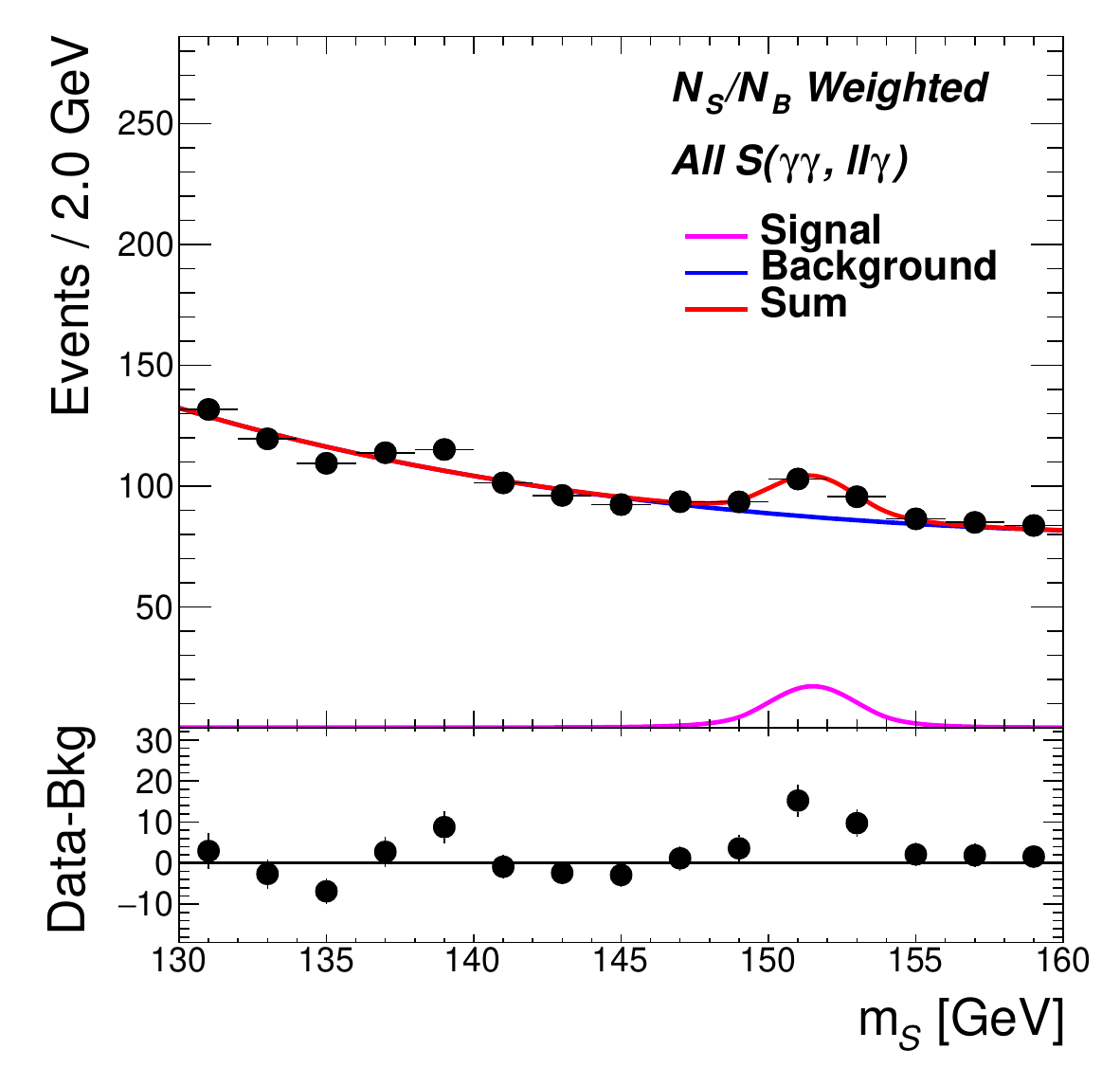}
            \caption{
 Left: The combined local $p$-value as a function of $m_S$ using inputs from Refs.~\cite{Sirunyan:2021ybb,ATLAS:2020pvn,Aad:2020ivc,Sirunyan:2020sum,Aad:2021qks,CMS:2018nlv,Sirunyan:2018tbk,HIGG-2018-51web} (see text).
Right: The combination of fits to all the categories from Refs.~\cite{Sirunyan:2021ybb,ATLAS:2020pvn,Aad:2020ivc,Sirunyan:2020sum,Aad:2021qks,CMS:2018nlv,Sirunyan:2018tbk,HIGG-2018-51web} (see text). The data points are $N_{S}/N_{B}$ weighted.}\label{fig:invmass2}
\end{figure*}

The spectra of the channels in which the excesses are most significant are shown in Fig.~\ref{fig:invmass1} and the combination of all channels in Fig.~\ref{fig:invmass2}, excluding the $b\overline{b}$ channel due insufficient resolution. Note that the background curves here do not exactly correspond to the SM hypothesis but rather to the background function of the combined fit to data (see appendix for fits showing both the original background obtained within the SM hypothesis and the refitted one including a NP resonance). For improved visualization, each plot combines the spectra from ATLAS and CMS (if available) for the same final states by using a signal over background ($N_{S}/N_{B}$) re-weighting, such that $N_{S}$ is the number of signal at the peak ($\pm 3\,$GeV) of the Crystal Ball function used to model the resonance and $N_B$ are the corresponding background events within this range.\footnote{The combination includes a rescaling of the luminosity and adjusting the efficiencies as well as the resolutions for ATLAS and CMS in case analyses of the same final state. This information is available in papers from both collaborations.}

We observe that the di-photon final states, in particular in channels with associated $E^{T}_{miss}$, contribute most to the excess. Furthermore, while there is a preference for a decay of $S$ to $Z\gamma$ and $b\overline{b}$ the significance for the latter ($1.2\sigma$) is quite small. We remark that the limits on the total cross section in the four-lepton final states (extracted from Refs.~\cite{HIGG-2016-19,CMS:2019chr}) show no hint for $S\to ZZ$. Note that the results are quite independent of the production mechanism of the di-photon resonance $S$, but that extractions of the total cross-sections from the fiducial ones would be model dependent, especially in the case of $E^{T}_{miss}$ searches. The total cross section can be estimated from the fiducial one by using the acceptance and efficiencies of the various channels in Refs.~\cite{CMS:2018nlv,Aad:2021qks}. As these quantities vary by $\approx$50\%, our total signal cross section should be about a factor $\approx 2$ times the fiducial one. Here we implicitly assume that the signal is generated together with $E^{T}_{miss}$, and that the latter is within the cuts that were applied in the analysis.

Let us now consider a simplified model in which $S$ is pair-produced by the decays of $H$ with a mass around 270\,GeV (as suggested by the multi lepton anomalies) via $pp\to H\rightarrow S S^{*}$. We remark that the process $H\rightarrow S h^{(*)}$ is possible in principle, but we will disregard it for now. The properties of $S$ can be determined approximately from the relative signal strengths of the considered channels: It should decay to photons, to a lesser extent to $Z\gamma$, and it should decay into invisible final states to produce missing energy. $S$ can further decay to $W$ bosons and bottom quarks with relevant branching ratios which is consistent with the assumptions that the couplings of $S$ to SM particles are induced via the mixing with $h$, i.e. that $S$ is SM-like. However, as $\gamma\gamma$ is only generated at the loop-level e.g.~by vector-like fermions (see e.g.~Refs.~\cite{Franceschini:2015kwy,Falkowski:2015swt,Benbrik:2015fyz,Chao:2015ttq}), it is reasonable to assume the corresponding width is modified by virtual new particles, which do not significantly affect $S\to WW$ or $S\to b \bar b$. Therefore, we assume that $H$ is produced (predominantly via gluon fusion) with a cross-section $\sigma_H$ and decays to approximately 100\% into $SS^*$. The scalar $S$, with a mass of $\approx 151\,$GeV, decays into di-photons with a small branching ratio ${\rm Br}(S\to \gamma\gamma) =\epsilon \ll 1$. In addition, it has a sizable branching ratio to $WW$ and $b\bar b$, as fixed by the requirement that it is SM-like, and otherwise decays into invisible final states $\chi$ with ${\rm Br}(S\to \chi\chi) = x=1-{\rm Br}(S\to WW)-{\rm Br}(S\to b\bar b)$.

In this setup, the signal strengths for each channel can be calculated as a function of two parameters, namely the total di-photon cross-section $\sigma_H \cdot \epsilon$ and the total cross-section to invisible. Thus we have two degrees of freedom and a simultaneous fit. Then, the maximal local significance for a resonance is 4.3$\sigma$ based on two degrees of freedom at $m_S=151.5\,$GeV, where it varies by 0.6$\sigma$ when shifting the mass by $1$\,GeV. Taking into account the look-else-where effect due to the mass scan between 140\,GeV to 155\,GeV, the significance is reduced to 3.9$\sigma$ with a trial factor $5$. The trial factor is obtained using the scan range divided by twice of the resolution 1.5\,GeV.\footnote{The spurious signal effect could degrade the significance by at most $\approx0.2$. Note that we that there is a 2.3$\sigma$ (local) excess reported by ATLAS in the search for fully hadronic final states~\cite{ATLAS:2020fgc} that can be interpreted as $S\rightarrow VV$ with $VV\to $ hadrons. This excess was not included in our fit since the spectrum was not given in a form usable for our analysis.}

\section{Asymmetric Di-Higgs searches}

The signatures we are discussing in this article, as well as the multi-lepton anomalies, can be explained by the decay of a neutral scalar $H$ into a lighter one $S$ and the SM Higgs~\cite{vonBuddenbrock:2016rmr,vonBuddenbrock:2018xar}, i.e. $H\rightarrow Sh,SS$, as realized e.g.\ within the 2HDM+$S$ model (also called N2HDM)~\cite{He:2008qm,Grzadkowski:2009iz,Chen:2013jvg,Muhlleitner:2016mzt,Krause:2017mal,Baum:2018zhf}.\footnote{Interestingly, the model can explain anomalies in astrophysics (the positron excess of AMS-02~\cite{PhysRevLett.122.041102} and the excess in gamma-ray fluxes from the galactic centre measured by Fermi-LAT~\cite{Ackermann_2017}) if it is supplemented by a Dark Matter candidate~\cite{Hektor:2015zba,Beck:2021xsv}.Furthermore, it can be easily extended~\cite{Sabatta:2019nfg} to account for the $4.2\sigma$ anomaly $g-2$ of the muon~\cite{Muong-2:2021ojo,Aoyama:2020ynm}.}

In order to verify or falsify the hypothesis of sizable $S\to b\overline{b}$ or $H\to Sh$ rates, one could search for $H\rightarrow SS\rightarrow\gamma\gamma b\overline{b}$ and $H\rightarrow Sh\rightarrow\gamma\gamma b\overline{b}$ final states. These are very promising signatures as they have the highest sensitivity for di-Higgs searches due to a good balance between the di-photon triggering efficiency, the triggering of the invariant mass spectra~\cite{CMS:2018tla,ATLAS:2018dpp}.

We illustrate this for $H\rightarrow SS\rightarrow\gamma\gamma b\overline{b}$. Assuming $m_H=270\,$GeV, the dominant branching ratio being $H\to SS^*$ forces one of the singlet scalars to be off-shell.\footnote{We notice that in other possible decay channels like $H\to S h$ and $H \to h h$ the daughter particles are on-shell, which offers different search strategies in the final states.} This type of resonant $\gamma\gamma b\overline{b}$ searches have not been performed by the LHC experiments. Here, two corners of the phase-space are devised to study asymmetric configurations: $m_{\gamma\gamma}\in(145,155)$\,GeV and $m_{b\overline{b}}\in(70,120)$\,GeV to isolate $H\rightarrow S(\rightarrow \gamma\gamma) S^*(\rightarrow b\overline{b})$; $m_{\gamma\gamma}\in(90,120)$\,GeV and $m_{b\overline{b}}\in(120,160)$\,GeV to isolate $H\rightarrow S(\rightarrow b\overline{b}) S^*(\rightarrow\gamma\gamma)$. To predict the resulting signatures, we perform a Monte-Carlo simulation of $pp$ collisions at the LHC. The events corresponding to the signal and SM backgrounds are generated using {\tt Madgraph5}~\cite{Alwall:2014hca} with the {\tt NNPDF3.0} parton distribution functions~\cite{NNPDF:2014otw}. The UFO model files required for the Madgraph analysis have been obtained from {\tt FeynRules}~\cite{Alloul:2013bka} after a proper implementation of the model. Following this parton\-level analysis, the parton showering and hadronization are performed using {\tt Pythia}~\cite{Sjostrand:2006za}. We use {\tt Delphes}(v3)~\cite{deFavereau:2013fsa} for the corresponding detector level simulation after the showering/hadronization. The jet construction at this level has been performed using { \tt Fastjet}~\cite{Cacciari:2011ma} which involves the anti-$K_{T}$ jet algorithm  with radius $R = 0.5$, transverse momentum $p_T>20$\,GeV and pseudorapidity $\left|\eta\right|<2.5$. Photons are required to have $p_T>20$\,GeV and $\left|\eta\right|<2.5$

\begin{figure}[t]
  \centering
   \includegraphics[width=0.45\textwidth]{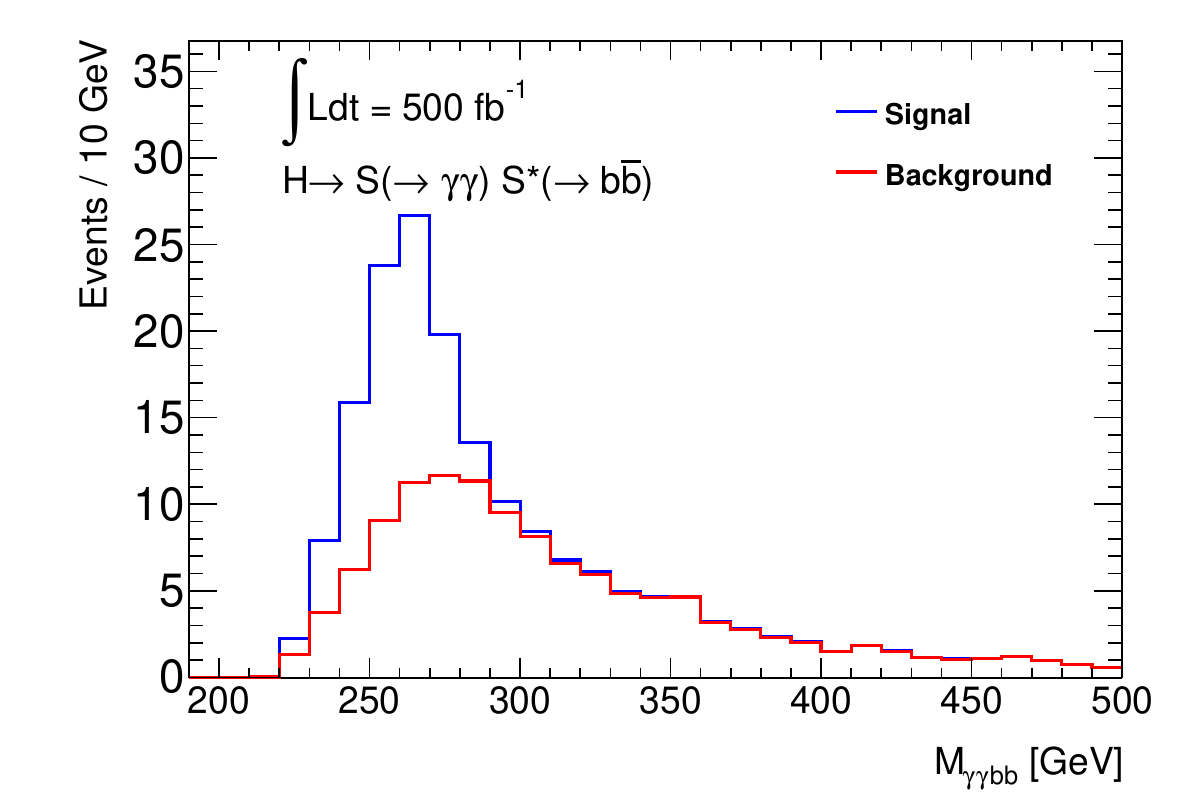}
   \includegraphics[width=0.45\textwidth]{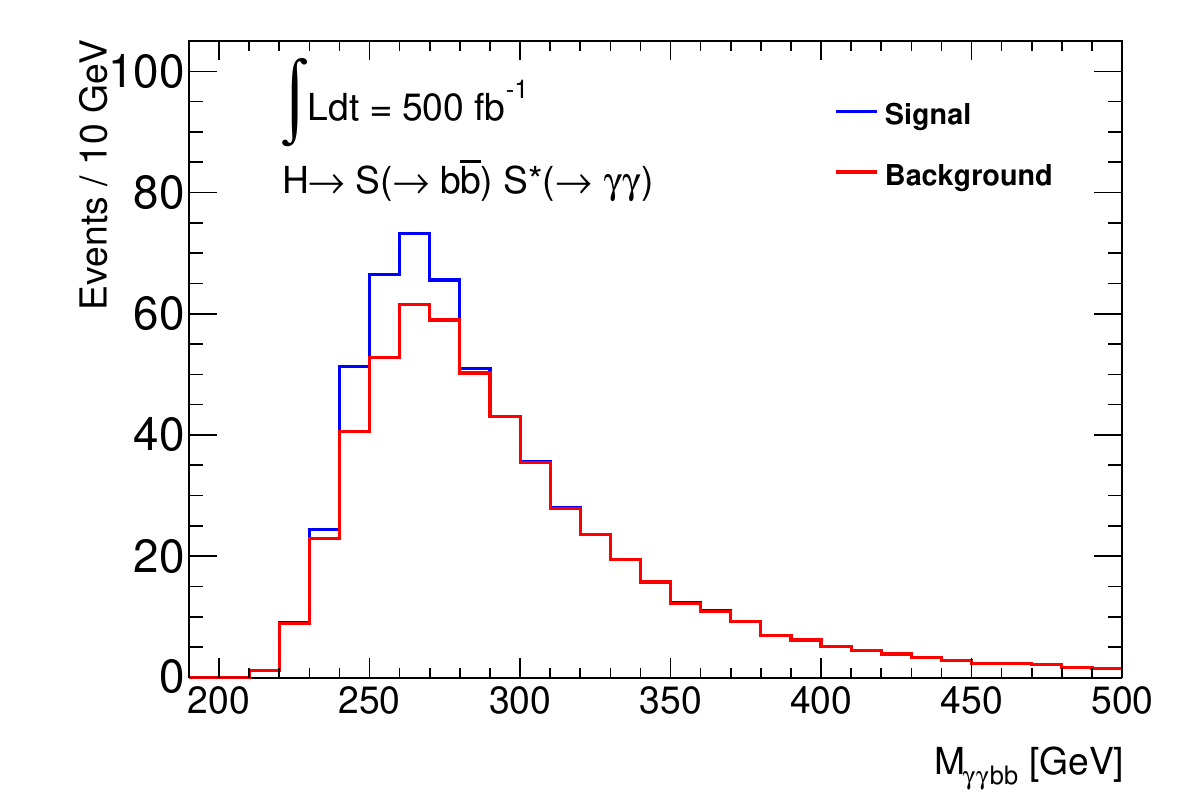}
 \caption{Expected yield of signal and background $\gamma\gamma b\overline{b}$ events for 500\,fb$^{-1}$ of integrated luminosity with one experiment at the LHC. The upper and and lower graphs correspond to the  $H\rightarrow S(\rightarrow \gamma\gamma) S^*(\rightarrow b\overline{b})$ and $H\rightarrow S(\rightarrow b\overline{b}) S^*(\rightarrow\gamma\gamma)$ searches, respectively. }
    \label{fig:yybb}
 \end{figure}

Figure~\ref{fig:yybb} shows the expected signal and background yields in $\gamma\gamma b\overline{b}$ final state for one LHC experiment and 500\,fb$^{-1}$ of integrated luminosity. The first configuration displays significantly better signal-to-background rates due to the excellent di-photon invariant mass resolution. Here the benchmark signal cross-section $\sigma(pp\rightarrow H\rightarrow SS^*\rightarrow\gamma\gamma b\overline{b})=2$\,fb is assumed, which includes the two configurations. The background includes the contribution from $\gamma\gamma$ in association with $c\overline{c}$ and light quarks. The $b$-jet tagging efficiency for $p_T=30$\,GeV is assumed to be 68\%. In this setup a combined significance of over 7$\sigma$ could be achieved per LHC experiment for 500\,fb$^{-1}$ of integrated luminosity, assuming the current best fit to data is confirmed. 

\section{Conclusions and Outlook}

In this article, we present evidence for an excess in the associate production that may provide evidence of a new Higgs particle $S$ (consistent with the assumption that it is produced from the decay of a heavier boson $H$) with a mass $m_S= 151.5$\,GeV by combining the CMS and ATLAS side bands ($130\,$GeV and $160\,$GeV) of searches for the SM Higgs. Including all channels involving photons, $Z$ bosons, light quarks, $b$-quarks and/or missing energy, we find a global significance of  $3.9\sigma$ (locally $4.3\sigma$) for $S$. 

While the hints that this new scalar is decaying into photons (and to a lesser extent into the $Z\gamma$) are significant, its decay to, or production together with, bottom quarks is preferred but still optional. To test the decay channel $S\to b\overline{b}$, we suggest performing asymmetric searches for the final state $\gamma\gamma b\overline{b},\tau^+\tau^- b\overline{b}$, which also allows an assessment of the branching ratio ${\rm Br}[H\to Sh]$. Furthermore, such signals could be related to the LEP/CMS excess in associate $b\overline{b}$ production and point towards a boson $S^\prime$ at $\approx 96\,$GeV that could lead to $H\to S^{(\prime)} S^\prime$. Furthermore, in Fig.~\ref{fig:invmass2} one can see a slight surplus of events around 138\,GeV which motivates further investigation. In fact, fitting this with an additional Crystal Ball function would increase the significance of the 151\,GeV excess even further. Together, this might point towards a whole undiscovered scalar sector.

Interestingly, as the LEP experiments reported a mild excess (2.3$\sigma$) in the search for a scalar boson ($S^\prime$)~\cite{LEPWorkingGroupforHiggsbosonsearches:2003ing} using the process $e^+e^-\rightarrow Zh(\rightarrow b\overline{b})$ at 98\,GeV for the invariant $b\overline{b}$ mass, asymmetric  $\gamma\gamma b\overline{b}$ final states could also originate from the decay $H\to SS^\prime$. This is further supported by the CMS result reporting similar excesses with Run 1 data and 35.9\,fb$^{-1}$ of Run 2 data~\cite{CMS:2018cyk}, with a local significance of 2.8$\sigma$ at 95.3\,GeV. In this context, it should be noted that searches with asymmetric configurations of  $H\rightarrow \tau^+\tau^- b\overline{b}$ are also substantiated.

\begin{figure*}[htbp!]
  \centering
  {\includegraphics[width=0.40\textwidth]{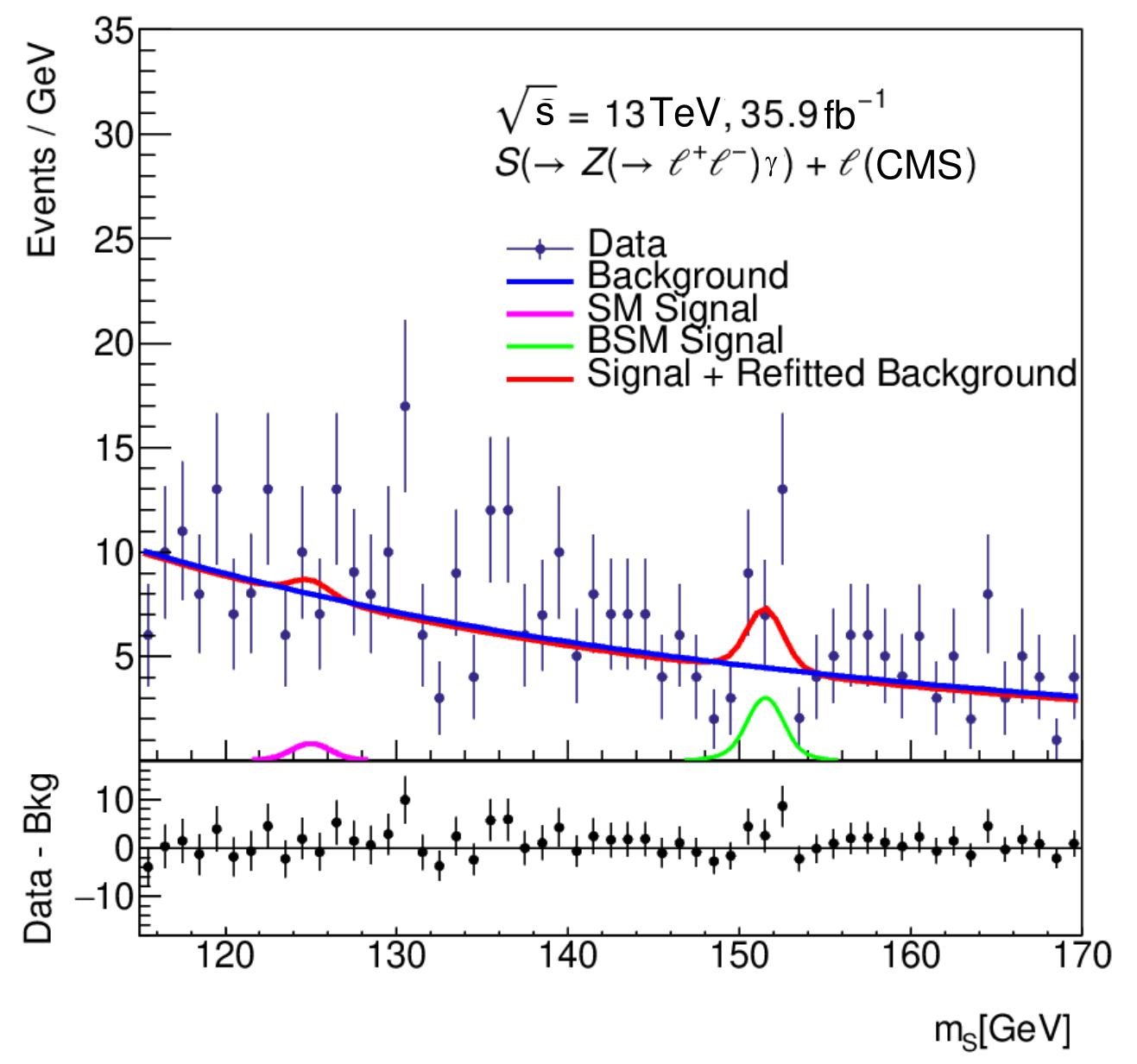}
  \label{fig:fig5a}}~~
  {\includegraphics[width=0.40\textwidth]{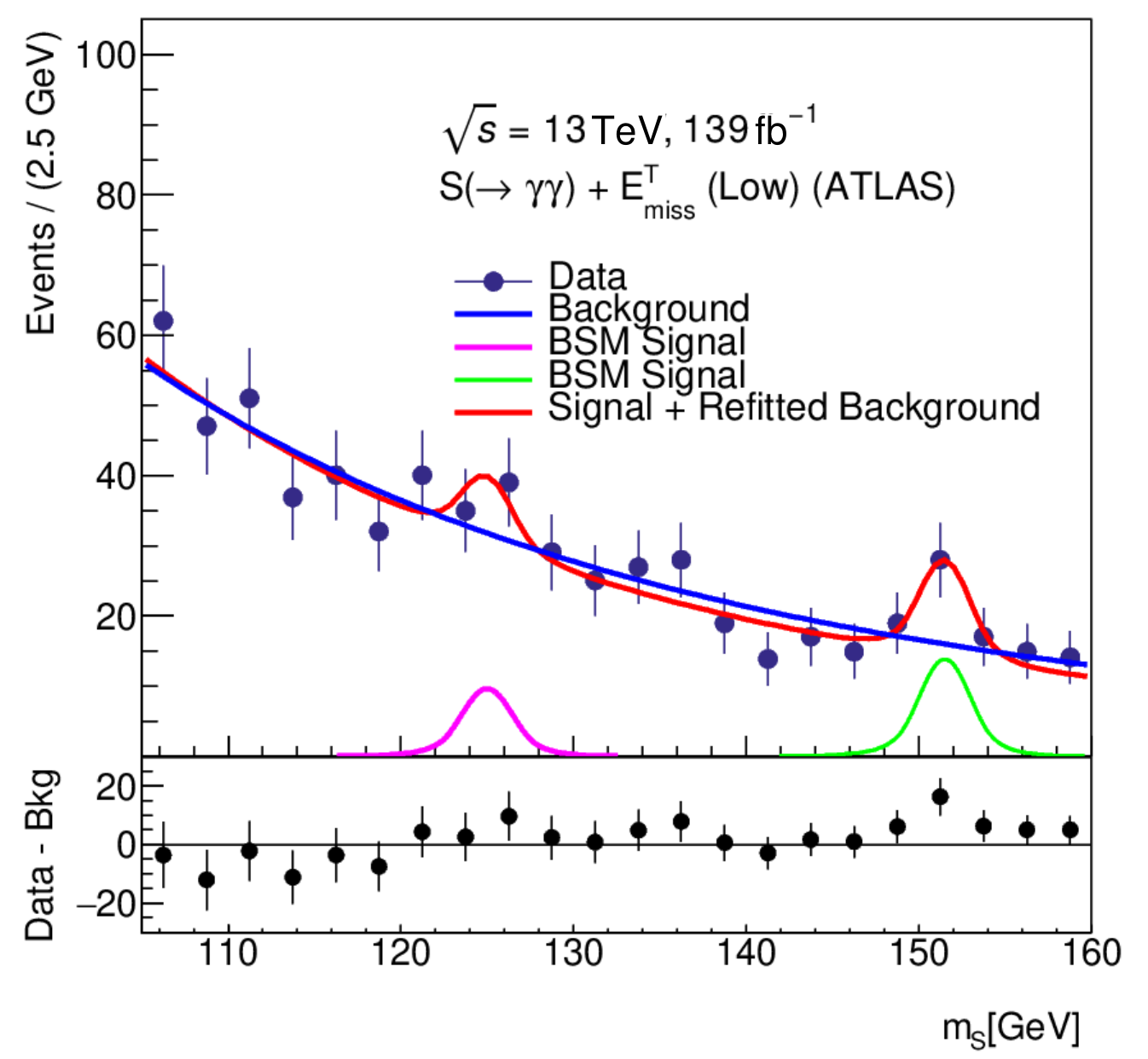}
  \label{fig:fig5b}}\\
  {\includegraphics[width=0.40\textwidth]{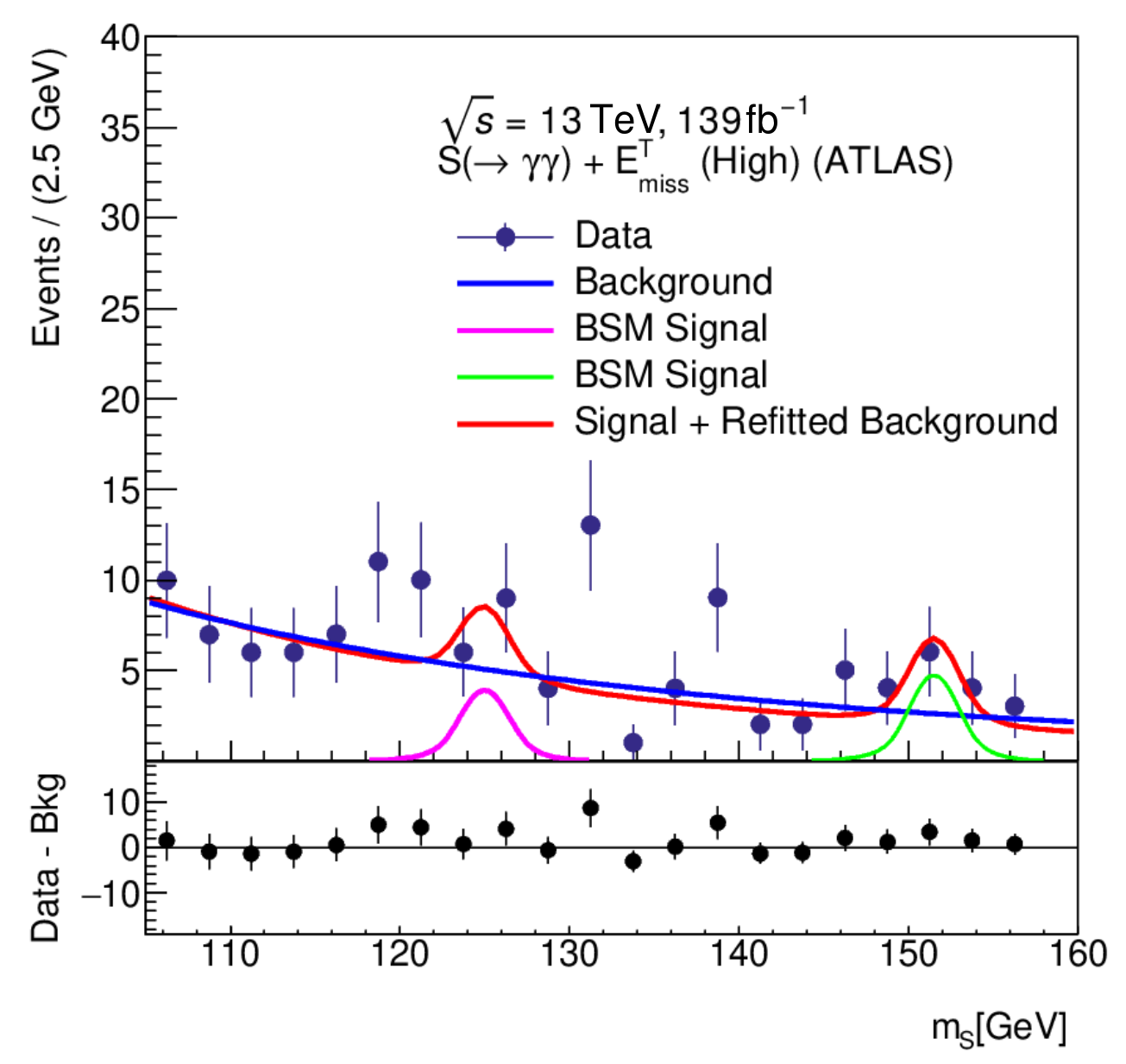}
  \label{fig:fig5c}}~~
  {\includegraphics[width=0.40\textwidth]{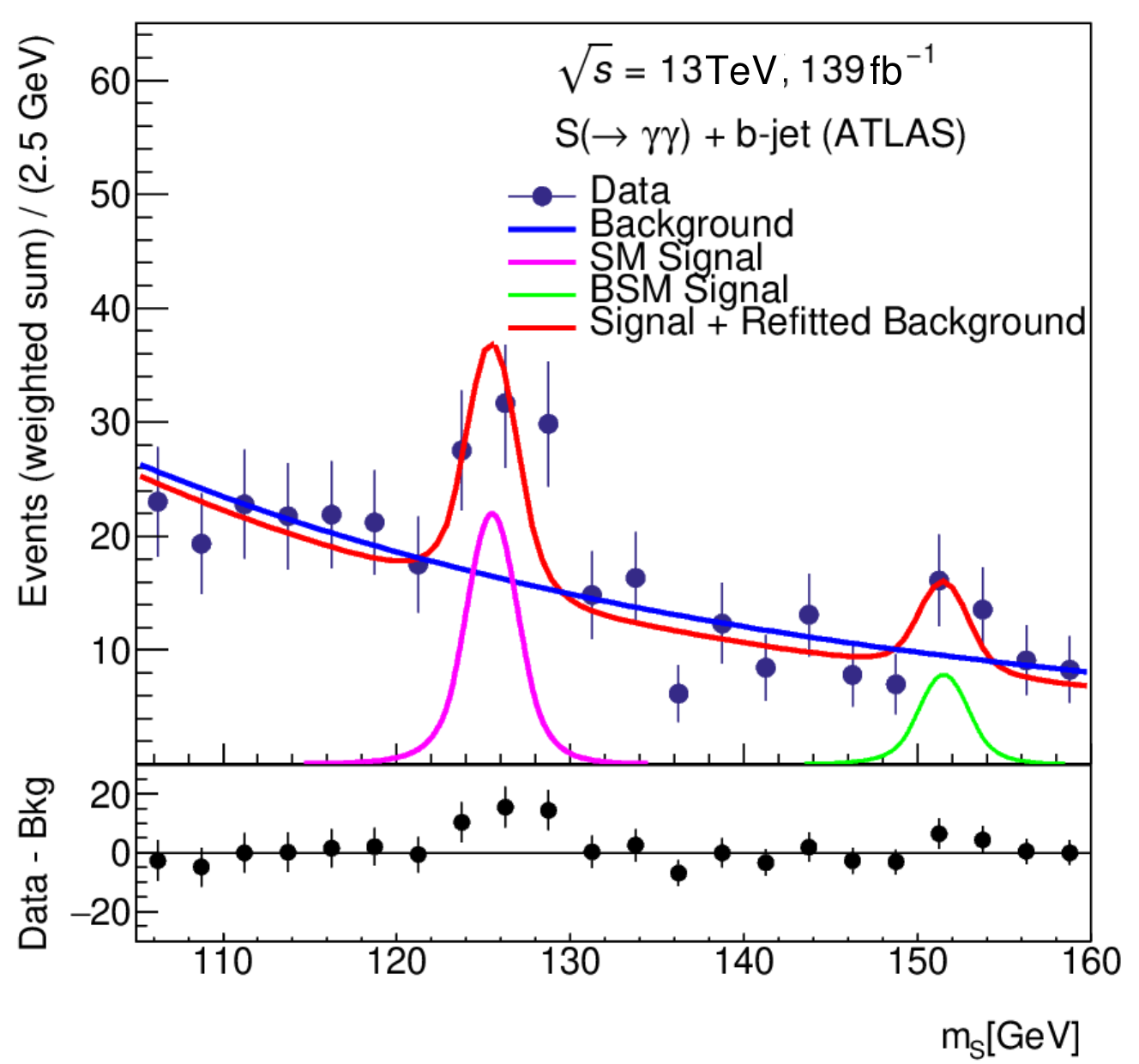}
  \label{fig:fig5d}}\\
  {\includegraphics[width=0.40\textwidth]{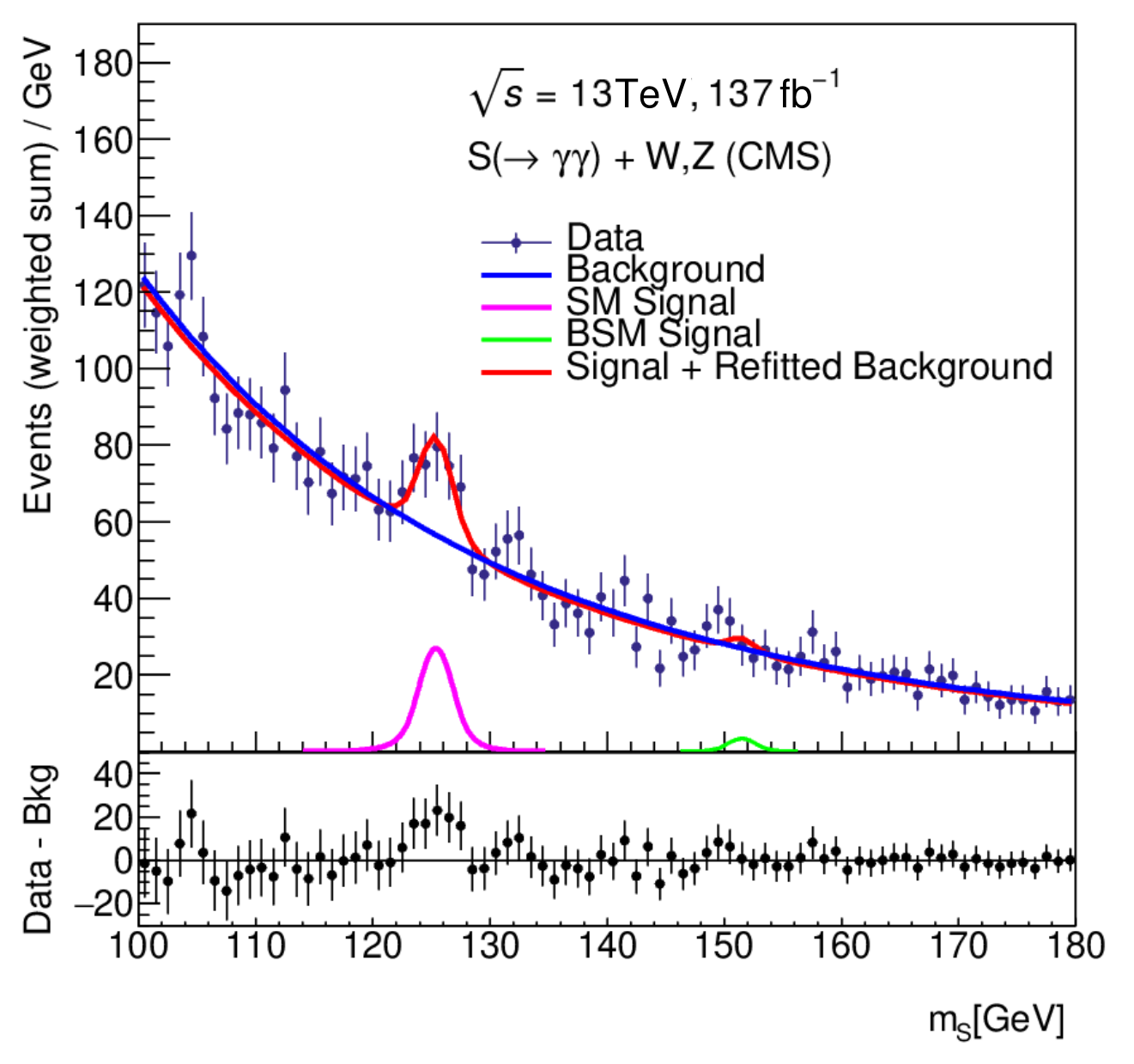}
  \label{fig:fig5e}}~~
  {\includegraphics[width=0.40\textwidth]{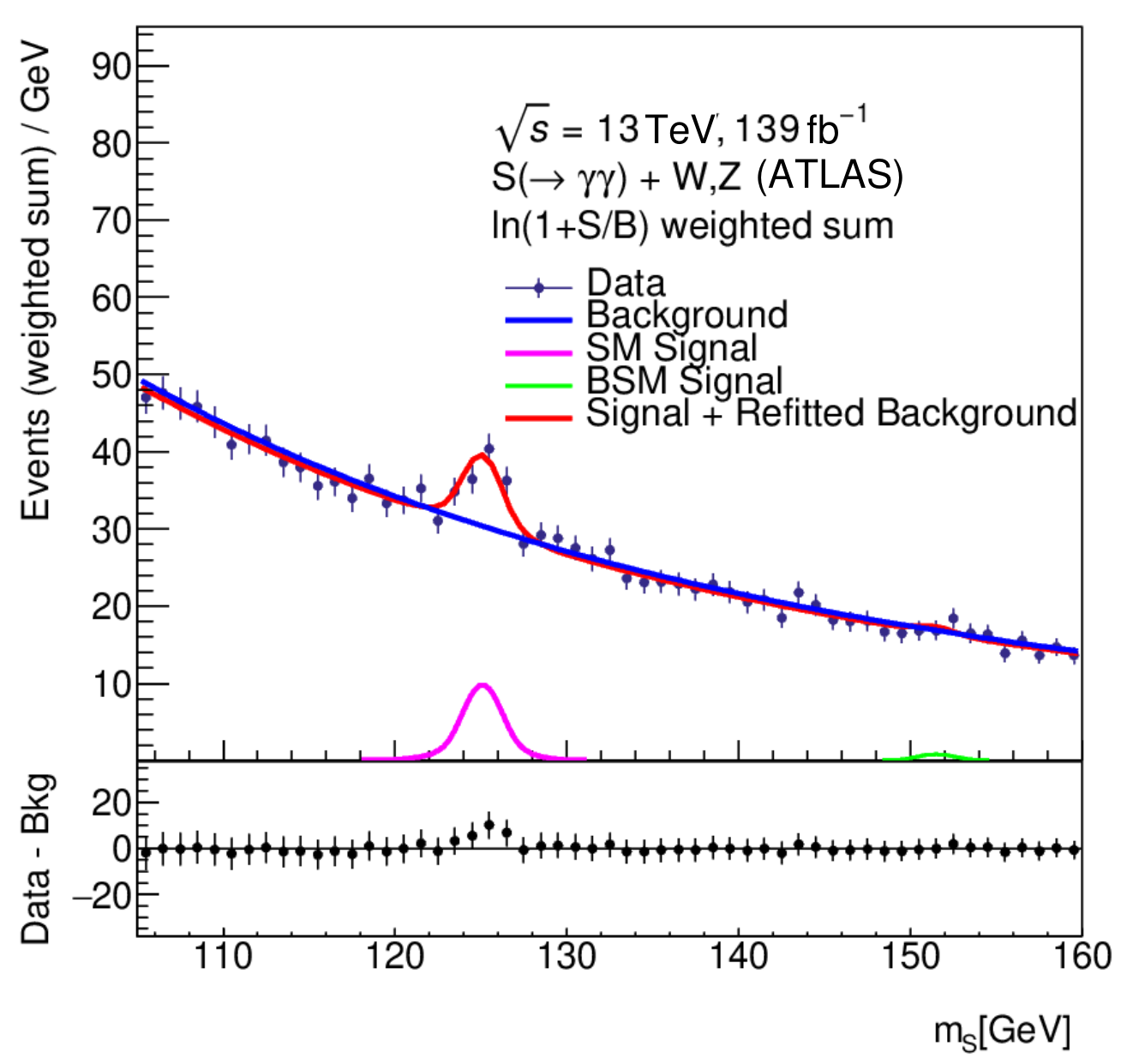}
  \label{fig:fig5f}}
  \caption{Diagrams showing the fit to background obtained within the SM, the SM Higgs signal and the NP signal with the refitted background for six different categories. The data displayed the last three plots corresponds to a weighted sum (see Refs.~\cite{Aad:2020ivc,Sirunyan:2021ybb,ATLAS:2020pvn} for further details).}
  \label{fig:fits}
 \end{figure*}

Our analysis potentially opens up new directions in particle physics. First of all, a mass of $S$ of 151\,GeV is motivated by the multi-lepton anomalies and therefore can obviously be related to them. However, due to the absence of a signal in $S\to ZZ^*$, an alternative mechanism for the lepton production, such as $S\to NN$, where $N$ has the quantum numbers of a right-handed neutrino, is possible. In such a setup, i.e.~the 2HDM+$S$ extended with right-handed neutrinos, the anomalous muon $g-2$ could be explained via a chiral enhancement~\cite{Crivellin:2018qmi,Chun:2020uzw,Ferreira:2021gke} and in case of a non-minimal flavour structure, also the intriguing indications for NP in $b\to s\ell^+\ell^-$ (above the $7\,\sigma$ level~\cite{Alguero:2021anc,Altmannshofer:2021qrr,Alok:2020bia,Hurth:2020ehu,Ciuchini:2020gvn}) could be explained~\cite{Li:2018rax,Marzo:2019ldg,Iguro:2018qzf,Crivellin:2019dun}. Furthermore, as $S$ is produced in association with missing energy, this opens new avenues for possible solutions to the outstanding problem of Dark Matter.

\begin{acknowledgments}
The authors are grateful for support from the South African Department of Science and Innovation through the SA-CERN program and the National Research Foundation for various forms of support. The work of A.C. is supported by a professorship grant from the Swiss National Science Foundation (No.\ PP00P21\_76884).
\end{acknowledgments}

\appendix

\section{Details on the individual fits}

{Figure~\ref{fig:fits} shows individual fits to ATLAS and CMS data for six different categories. The background obtained within the SM, the SM Higgs signal and the NP signal with the refitted background are shown.}

\bibliography{apssamp}

\end{document}